\documentclass[fleqn,usenatbib]{mnras}
\usepackage{newtxtext,newtxmath}
\usepackage[T1]{fontenc}
\usepackage{ae,aecompl}
\usepackage{graphicx}
\usepackage{dcolumn}
\usepackage{bm}
\usepackage{color}
\usepackage{natbib}
\usepackage{ulem}
\usepackage{booktabs}

\newcommand{\appropto}{\mathrel{\vcenter{
  \offinterlineskip\halign{\hfil$##$\cr
    \propto\cr\noalign{\kern2pt}\sim\cr\noalign{\kern-2pt}}}}}

\newcommand{\vs}{v_{\mathrm{s}}}

\newcommand{\rmax}{r_{\mathrm{max}}}
\newcommand{\mvir}{M_{\mathrm{vir}}}
\newcommand{\rvir}{R_{\mathrm{vir}}}

\newcommand{\rs}{r_{\mathrm{s}}}

\newcommand{\rhos}{\rho_{\mathrm{s}}}

\newcommand{\msunh}{h^{-1} M_{\odot}}

\newcommand{\kpch}{h^{-1} \mathrm{kpc}}

\title[Scaling Relations of Dark Matter Haloes]{Scaling Relations in the Phase-Space Structure of Dark Matter Haloes
}
\author[Gross et al.]{
Axel Gross,$^{1}$
Zhaozhou Li,$^{2}$\thanks{E-mail: \href{mailto:lizz.astro@gmail.com}{lizz.astro@gmail.com}}
and Yong-Zhong Qian$^{1}$\\
$^{1}$School of Physics and Astronomy, University of Minnesota, Minneapolis, MN 55455, USA\\
$^{2}$Centre for Astrophysics and Planetary Science, Racah Institute of Physics, 
      The Hebrew University, Jerusalem, 91904, Israel}

\pubyear{2024}
\begin{document}
\label{firstpage}
\pagerange{\pageref{firstpage}--\pageref{lastpage}}
\maketitle
\begin{abstract}
We present new scaling relations for the isotropic phase-space distribution functions (DFs) and energy distributions of simulated dark matter haloes. These relations are inspired by those for the singular isothermal sphere with density profile $\rho(r)\propto r^{-2}$, for which the DF satisfies $f(E) \propto \rmax^{-2}(E)$ and the energy distribution satisfies $dM/dE \propto \rmax(E)$, with $\rmax(E)$ being the radius where the gravitational potential equals energy $E$. For the simulated haloes, we find $f(E)\propto\rmax^{-2.08}(E)$ and $dM/dE \propto \rmax(E)$ across broad energy ranges. In addition, the proportionality coefficients depend on the gravitational constant and the parameters of the best-fit Navarro-Frenk-White density profile. These scaling relations are satisfied by haloes over a wide mass range and provide an efficient method to approximate their DFs and energy distributions. Understanding the origin of these relations may shed more light on halo formation.
\end{abstract}

\section{Introduction}
For a dark matter halo in dynamical equilibrium, its distribution function (DF) $f(\bm{r},\bm{v})=d^6M/d^3\bm{r}d^3\bm{v}$ in the phase space of particle position $\bm{r}$ and velocity $\bm{v}$ provides a complete description. However, cosmological N-body simulations usually do not have the requisite number of particles to properly sample the 6-dimensional phase space (cf. \citealt{2006MNRAS.373.1293S}). Furthermore, it can be challenging to self-consistently obtain the DF from quantities such as the energy distribution and profiles of density and velocity anisotropy, which are more easily determined from simulations. Consequently, despite many attempts (e.g., \citealt{1991MNRAS.253..414C,PhysRevD.73.023524,Wojtak2008,10.1093/mnras/stu2608,10.1093/mnras/stv096}), an accurate and complete form of the DF for dark matter haloes remains to be found. 

The energy distribution $dM/dE$ is known to be mostly dependent on the density profile of the halo, with a weak dependence on the velocity anisotropy \citep{2008gady.book.....B,2021A&A...653A.140B}. Many studies have focused on the isotropic case, where the DF is a function of energy only, $f(\bm{r},\bm{v})=f(E)=(dM/dE)/g(E)$ with $g(E)$ being the density of energy states. Although the isotropic DF can be determined from the density profile via the Eddington inversion \citep{10.1093/mnras/76.7.572}, the result for the commonly-used Navarro-Frenk-White (NFW, \citealt{1997ApJ...490..493N}) profile is not analytical and must be derived numerically or fitted to some complicated form \citep{2000ApJS..131...39W}. In this paper, we present new scaling relations that provide an efficient method to approximate the isotropic DFs and energy distributions of dark matter haloes.

Other scaling relations for simulated haloes have been reported in the literature. Notably, as first presented by \cite{2001ApJ...563..483T} and subsequently explored by many others (e.g., \citealt{2005ApJ...634..756A}; \citealt{2007ApJ...671.1108H}; \citealt{2011MNRAS.415.3895L}; \citealt{2020ApJ...893...53A}), the pseudo-phase-space density follows a power law, $\rho(r)/\sigma^3(r) \propto r^{-\gamma}$ with $\gamma=1.875$, where $\rho(r)$ and $\sigma(r)$ are the density and velocity dispersion at radius $r$, respectively. In addition, the volumetric density of the DF, ${\cal{V}}(f_0)=\int d^3\bm{r}d^3\bm{v}\delta(f(\bm{r},\bm{v})-f_0)$, was shown to approximately follow the power law ${\cal{V}}(f)\propto f^{-2.5}$ over a wide range of $f$ values \citep{10.1111/j.1365-2966.2004.08045.x}. We will show that this result on ${\cal{V}}(f)$ can be accounted for by our scaling relations.

We outline our paper as follows. In \S\ref{sec:background}, we review the formalism of the DF and its connections to the density profile and energy distribution. We illustrate this formalism by deriving the scaling relations for the DF and energy distribution of the singular isothermal sphere (SIS) with $\rho(r)\propto r^{-2}$. In \S\ref{sec:halo}, we describe the sample of simulated haloes and present the scaling of the median energy distribution and DF with the radius $\rmax(E)$ at which the gravitational potential equals the energy $E$. In \S\ref{sec:use}, we show that these scaling relations provide an empirical model to approximate the DFs and energy distributions of individual simulated haloes. We find that this empirical model has comparable accuracy to the DARKexp fit of \cite{2010ApJ...722..851H}. We also outline procedures to estimate the DF and energy distribution of a halo based on its best-fit NFW profile. In \S\ref{sec:origin}, we show that our scaling relations can account for the result of \cite{10.1111/j.1365-2966.2004.08045.x} on ${\cal{V}}(f)$, and qualitatively discuss the origin of the scaling relation for the energy distribution. In \S\ref{sec:conclude}, we summarize our results and give conclusions. 

\section{DF Formalism and SIS Scaling Relations}
\label{sec:background}
We focus on haloes with spherical density profiles and isotropic velocity distributions, for which the DF is a function of the energy per unit mass $E=v^2/2+\Phi(r)$ only, where $v$ is the velocity and $\Phi(r)$ is the gravitational potential. We define the DF as a mass distribution in phase space:
\begin{align}
\label{eq:feconvention}
    f(E)=\frac{d^6M}{d^3\bm{r}d^3\bm{v}}=m_{\rm d}\frac{d^6N}{d^3\bm{r}d^3\bm{v}},
\end{align}
where $m_{\rm d}$ is the mass of the dark matter particle. The corresponding density profile is given by
\begin{align}
\label{eq:densityfromfe}
    \rho(r)=\int d^3\bm{v}\,f(E)=4\pi\int_{\Phi(r)}^{\Phi(\infty)} dE\,f(E)\sqrt{2[E-\Phi(r)]},
\end{align}
which can be inverted to give \citep{10.1093/mnras/76.7.572}
\begin{align}
\label{eq:eddington}
    f(E)=\frac{1}{\pi^2\sqrt{8}}\frac{d}{dE}\int_{\rmax}^{\infty}
    \frac{dr}{\sqrt{\Phi(r)-E}}\frac{d{\rho}}{dr}.
\end{align}  
For consistency, $\rho(r)$ and $\Phi(r)$ must also satisfy Poisson's equation $\nabla^2\Phi(r)=4\pi G\rho(r)$, where $G$ is the gravitational constant. In Eq.~(\ref{eq:eddington}), $\rmax$ corresponds to $\Phi(\rmax)=E$ and is the key quantity for our scaling relations. It will be written as $\rmax(E)$ in these relations to emphasize its dependence on $E$. Mathematically, $\rmax(E)$ is the inverse function of $\Phi(r)$. 

The energy distribution $dM/dE$ is related to the DF by
\begin{equation}
    \frac{dM}{dE}=f(E)g(E),
\label{eq:dmde}
\end{equation}
where
\begin{equation}
\begin{split}
    g(E)&=\int d^3\bm{r} d^3\bm{v}\,\delta \left(\frac{v^2}{2}+\Phi(r)-E\right)\\
    &=16\pi^2\int_0^{\rmax} dr\,r^2 \sqrt{2[E-\Phi(r)]}
\label{eq:ge}
\end{split}
\end{equation}
is the density of energy states. 

The above formalism gives a set of self-consistent descriptions of a halo in terms of $\rho(r)$, $\Phi(r)$, $f(E)$, and $dM/dE$. In principle, any of the four can be used to determine the others. As a concrete example, we apply this formalism to the SIS with the density profile
\begin{equation}
\rho(r)= \frac{\sigma_{\rm SIS}^2}{2\pi G r^2},
\end{equation}
where $\sigma_{\rm SIS}$ is the constant velocity dispersion at any radius [see Eq.~(\ref{eq:sigmasis})]. Note that $\rho(r)/\sigma^3(r) \propto r^{-2}$ for the SIS is a special case of the power law for pseudo-phase-space density \citep[cf.][]{2001ApJ...563..483T}. The SIS gravitational potential is given by 
\begin{equation}
    \Phi(r)=2\sigma_{\rm SIS}^2 \ln(r/r_0),
\end{equation}
with $\Phi(r_0)=0$. Inverting $\Phi(r)$, we obtain
\begin{equation}
    r=r_0\exp[\Phi/(2\sigma_{\rm SIS}^2)],
\end{equation}
which gives
\begin{align}
    \rmax(E) &= r_0\exp[E/(2\sigma_{\rm SIS}^2)],\label{eq:rmaxsis}\\
    \rho(r) &= \frac{\sigma_{\rm SIS}^2}{2\pi Gr_0^2}\exp(-\Phi/\sigma_{\rm SIS}^2).\label{eq:rhophi}
\end{align}
Using Eq.~(\ref{eq:eddington}) along with Eqs.~(\ref{eq:rmaxsis}) and (\ref{eq:rhophi}), we obtain
\begin{align}
    f(E)&= \frac{1}{(2\pi)^{5/2}G\sigma_{\rm SIS} r_0^2}\exp(-E/\sigma_{\rm SIS}^2)\label{eq:sigmasis}\\
    &= \frac{1}{(2\pi)^{5/2}G\sigma_{\rm SIS}}\rmax^{-2}(E).\label{eq:fesis}
\end{align}  
Similarly, we obtain from Eq.~(\ref{eq:ge})
\begin{align}
\label{eq:gesis}
g(E)&=\frac{16\pi^{5/2}\sigma_{\rm SIS} r_0^3}{3^{3/2}}\exp[3E/(2\sigma_{\rm SIS}^2)]\nonumber\\
&= \frac{16\pi^{5/2}\sigma_{\rm SIS}}{3^{3/2}}\rmax^{3}(E),
\end{align}
which along with Eq.~(\ref{eq:fesis}) gives
\begin{equation}
\label{eq:dmdesis}
\frac{dM}{dE}= \left(\frac{2}{3}\right)^{3/2}\frac{\rmax(E)}{G}.
\end{equation}

The above example demonstrates the exact scaling relations $f(E) \propto \rmax^{-2}(E)$ and $dM/dE \propto \rmax(E)$ for the SIS. The dependence on $G$ and $\sigma_{\rm SIS}$ for the proportionality coefficients of these relations can be deduced from dimensional analysis. While the choice of $r_0$, at which radius the potential vanishes, enters these relations through $\rmax(E)$ [see Eq.~(\ref{eq:rmaxsis})], it only affects the numerical value of $\rmax(E)$. Clearly, the power indices of the SIS scaling relations are independent of $r_0$. 

\section{Scaling Relations for Simulated Haloes}
\label{sec:halo}
Motivated by the SIS scaling relations, we now present similar scaling relations for simulated haloes. It is well known that the spherically-averaged density profile of dark matter haloes can be well fitted by the NFW profile \citep{1997ApJ...490..493N} between $0.05\rvir$ and $\rvir$ with $\rvir$ being the virial radius, though deviations can occur outside this range \citep{2015MNRAS.451.1247S}. The NFW profile is given by
\begin{equation}
\label{eq:nfwrho}
\rho_{\rm NFW}(r)=\frac{\rhos}{(r/\rs)(1+r/\rs)^2},
\end{equation}
where $\rhos$ and $\rs$ are the characteristic scales for density and radius, respectively. The corresponding gravitational potential is given by
\begin{align}
    \Phi_{\rm NFW}(r)=\vs^2\left[\frac{\ln(1+\rvir/\rs)}{\rvir/\rs}-\frac{\ln(1+r/\rs)}{r/\rs}\right],
\label{eq:nfwphi}
\end{align}
where $\vs=\rs\sqrt{4\pi G\rhos}$\,. For convenience of discussion below, we have chosen $\Phi_{\rm NFW}(\rvir)=0$. While we will use the numerical potential from simulations, the parameters of the best-fit NFW profile provide useful physical scales for a simulated halo.

We look for relations of the form
\begin{align}
\frac{dM}{dE}&=\frac{\alpha\rs}{G}\left[\frac{\rmax(E)}{\rs}\right]^m=\frac{\alpha\rvir}{G}c^{m-1}\left[\frac{\rmax(E)}{\rvir}\right]^m,\label{eq:dmdermax1}\\
f(E)&=\frac{\beta}{Gv_s r_s^{2}}\left[\frac{\rmax(E)}{\rs}\right]^n=\frac{\beta}{Gv_s\rvir^{2}}c^{n+2}\left[\frac{\rmax(E)}{\rvir}\right]^n,\label{eq:fermax1}
\end{align}
where $c=\rvir/\rs$ is the halo concentration, $\alpha$ and $\beta$ are dimensionless constants, and $m$ and $n$ are power indices. The NFW profile changes from $\rho\propto r^{-1}$ at $r\ll\rs$ to $\rho\propto r^{-3}$ at $r\gg\rs$, and behaves like the SIS at $r\sim\rs$. While the SIS scaling relations suggest $m\approx 1$ and $n\approx -2$, we will obtain the best-fit values for $m$ and $n$ along with those for $\alpha$ and $\beta$ from the data on simulated haloes. The fitting procedure will also determine the ranges of $\rmax(E)$ for which the scaling relations provide good description of the data.

\subsection{Halo Sample}
\label{sec:simdetail}
As in \cite{Gross:2024dsw}, we use the haloes in the TNG300-1-Dark simulations from the IllustrisTNG Project \citep{10.1093/mnras/stx3304,10.1093/mnras/stx3112,10.1093/mnras/stx3040,10.1093/mnras/sty618,2018MNRAS.480.5113M}. The simulations were performed with the Planck Collaboration XIII \citeyearpar{2016A&A...594A..13P} cosmological parameters: $\Omega_\mathrm{m} =  0.3089$, $\Omega_\mathrm{b} = 0.0486$, $\Omega_{\Lambda} = 0.6911$, $h=0.6774$, $n_\mathrm{s}=0.9667$, and $\sigma_8 = 0.8159$. The particle mass is $m_{\rm d}=7\times10^7\msunh$ and the softening length is $1\kpch$. The virial mass and radius of a halo, $\mvir$ and $\rvir$, respectively, are determined such that the average density inside $\rvir$ is equal to $\Delta_\mathrm{vir}$ times the critical density of the Universe, where $\Delta_\mathrm{vir}=18\pi^2+82(\Omega_\mathrm{m}-1)-39(\Omega_\mathrm{m}-1)^2\approx 102$ is the virial factor \citep{1998ApJ...495...80B}. 

We select a random sample of 100 isolated haloes in the range of $\mvir \in [10^{12},10^{14.5}]\msunh$, based on the criteria that an isolated halo does not contain any subhaloes of mass exceeding $0.1\mvir$ within $2.5\rvir$ of its vicinity, and that it does not overlap with the $2.5\rvir$ vicinity of any halo of mass exceeding $0.5\mvir$. We further select relaxed haloes, for which the deviation between the center of mass and the location of minimum potential is $\Delta r=|\bm{r}_\mathrm{CoM} - \bm{r}_\mathrm{MinPot}| < 0.01\rvir$. This relaxation criterion removes 19 haloes from our sample. For each of the remaining haloes, we fit the density distribution to the NFW profile by minimizing the root-mean-squared (rms) deviation of $\log\rho$ in the range of $[0.05, 1] \rvir$. We further remove two haloes that have very poor NFW fits with $(c, \mvir,\delta_{\log \rho})=(2.9,3.2\times10^{13} \msunh, 0.11)$ and $(4.4,4.6\times10^{13} \msunh, 0.10)$. The final sample contains 79 isolated and relaxed haloes.

For each halo of the final sample, we construct $dM/dE$ by counting the particles within $r_{\textrm{lim}}=2.5\rvir$, which is approximately the depletion radius separating a growing halo from its draining environment \citep{2021MNRAS.503.4250F,2023ApJ...953...37G}. As the energy is very sensitive to the choice of reference frame, we use the mean values for the most bound 10\% of the particles in a halo to define its position and velocity \citep[cf.][]{2012MNRAS.427.2437H}. The DF is obtained as $f(E)=(dM/dE)/g(E)$ (e.g., \citealt{1997MNRAS.286..329N}). Because we only count particles with $r<r_{\textrm{lim}}$, the corresponding $g(E)$ is
\begin{equation}
    g(E)=16\pi^2\int_0^{r^*_{E}} dr\,r^2 \sqrt{2[E-\Phi(r)]},
\label{eq:ge2}
\end{equation}
where $r^*_E=\text{min}(r_{\textrm{lim}},\rmax(E))$. These results on $dM/dE$ and $f(E)$ will be compared with the scaling relations and other approximations in \S\ref{sec:use} (see also \citealt{Gross:2024dsw}).

\subsection{Fitting of Scaling Relations}
\label{sec:simfit}
\begin{figure*}
\centering
\includegraphics[width=1\textwidth]{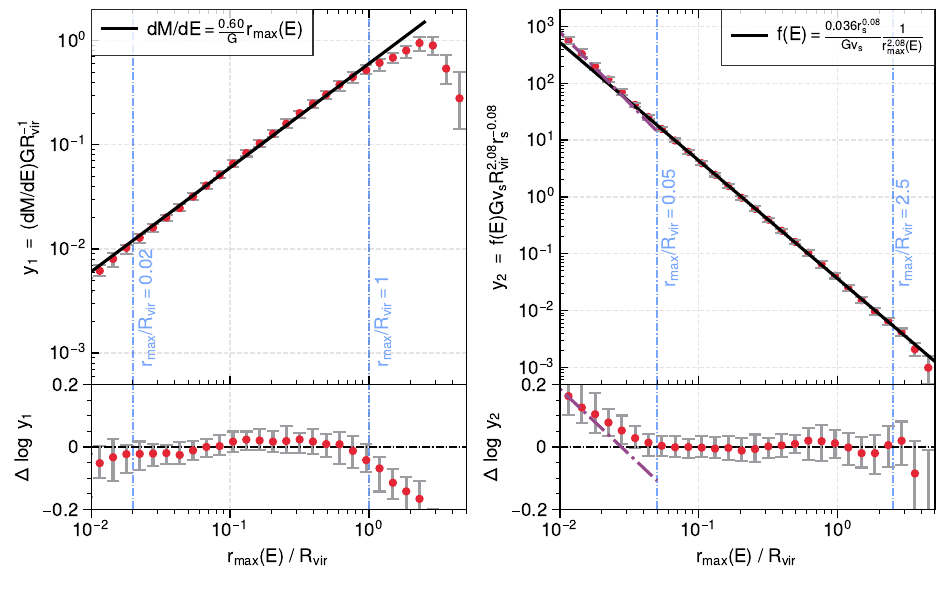}
\vspace{-2em}
\caption{%
\textit{Left panel}: The top part shows the median value of $y_1=(dM/dE)G/\rvir$ (red filled circles) and $1\sigma$ scatter (grey bars) as functions of binned $\rmax(E)/\rvir$ based on our sample of 79 haloes with $\mvir \in [10^{12},10^{14.5}]\msunh$. For comparison, the fitted scaling relation given in the inset is shown as the solid line. The bottom part shows the difference $\Delta\log y_1$ between the halo data and the solid line. The blue vertical lines bound the fitting range of $\rmax(E)\in [0.02\rvir, \rvir]$ for the scaling relation. \textit{Right panel}: Same as left panel, but for $y_2=f(E)Gv_s\rvir^{2.08}/\rs^{0.08}$. The blue vertical lines bound the fitting range of $\rmax(E)\in [0.05\rvir, 2.5\rvir]$ for the scaling relation in the inset. The purple dashed line shows the trend of $f(E) \propto \rmax^{-2.5}(E)$ at small radii. See text for details.
}
\label{fig:scaling}
\end{figure*}

Rather than fitting the scaling relation for $dM/dE$ in Eq.~(\ref{eq:dmdermax1}) to each halo, we aim to find a relation that applies to all the haloes. We obtain the numerical constant $\alpha$ and power index $m$ in this relation as follows. Because particles with $r\sim\rvir$ are of main concern, we use a uniform logarithmic grid of $\rmax(E)/\rvir$ for all the haloes. For each halo, we recalculate $dM/dE$ by counting the particles in each logarithmic bin of $\rmax(E)/\rvir$ (i.e., $E$ is an implicit variable relating $\rmax(E)$ to $dM/dE$). Note that over the fitting range below, there are at least 10 particles in each bin for every halo so that $dM/dE$ can be calculated to reasonable accuracy. For each bin, we sort values of $(dM/dE)Gc^{1-m}/\rvir$ for all the haloes to find the median. By inspection, the median value of $(dM/dE)Gc^{1-m}/\rvir$ as a function of $\rmax(E)/\rvir$ follows the scaling relation in Eq.~(\ref{eq:dmdermax1}) over $\rmax(E)\in[0.02\rvir,\rvir]$ (this range is chosen to achieve the optimal fit). Detailed fitting by minimizing the rms deviation of $\log[(dM/dE)Gc^{1-m}/\rvir]$ over this range gives $\alpha=0.60$ and $m=1.01$ with $\delta_{\log[(dM/dE)G\rs^{0.01}/\rvir^{1.01}]}=0.019$. Because the fitted $m$ value is so close to 1, we repeat the above procedure by setting $m=1$ and obtain $\alpha=0.60$ with $\delta_{\log[(dM/dE)G/\rvir]}=0.020$. In view of the simplicity without loss of accuracy, we adopt 
\begin{align}
    \frac{dM}{dE}=\frac{0.60}{G}\rmax(E),\label{eq:dmdermax2}
\end{align}
which differs from the SIS scaling relation [Eq.~(\ref{eq:dmdesis})] only in the numerical constant (0.54 for the SIS). In the left panel of Fig.~\ref{fig:scaling}, we show the median value of $(dM/dE)G/\rvir$ and the $1\sigma$ scatter as functions of the binned $\rmax(E)/\rvir$. For comparison, we also show the scaling relation in Eq.~(\ref{eq:dmdermax2}) as the solid line.

The procedure for fitting the scaling relation for $dM/dE$ can be repeated to fit that for $f(E)$. By inspection, the median value of $f(E)Gv_s\rvir^2/c^{2+n}$ as a function of $\rmax(E)/\rvir$ follows the scaling relation in Eq.~(\ref{eq:fermax1}) over $\rmax(E)\in[0.05\rvir,2.5\rvir]$. Detailed fitting by minimizing the rms deviation of $\log[f(E)Gv_s\rvir^2/c^{2+n}]$ over this range gives $\beta=0.036$ and $n=-2.08$ with $\delta_{\log[f(E)Gv_s\rvir^{2.08}/\rs^{0.08}]}=0.011$. By comparison, setting $n=-2$ as for the SIS gives $\beta=0.033$ with $\delta_{\log[f(E)Gv_s\rvir^2]}=0.041$. This $\beta$ value corresponds to the SIS with $\sigma_{\rm SIS}=\vs/3.3$ [see Eq.~(\ref{eq:fesis})]. Because the fit with $n=-2.08$ is significantly better, we adopt
\begin{equation}
\label{eq:fermax2}
f(E)=\frac{0.036r_s^{0.08}}{Gv_s}\frac{1}{\rmax^{2.08}(E)}.
\end{equation}
In the right panel of Fig.~\ref{fig:scaling}, we show the median value of $f(E)Gv_s\rvir^{2.08}/\rs^{0.08}$ and the $1\sigma$ scatter as functions of the binned $\rmax(E)/\rvir$. For comparison, we also show the scaling relation in Eq.~(\ref{eq:fermax2}) as the solid line.

It can be seen from Fig.~\ref{fig:scaling} that the scaling relations in Eqs.~(\ref{eq:dmdermax2}) and (\ref{eq:fermax2}) describe the halo data very well within the respective fitting ranges. For $dM/dE$, the halo data fall below the scaling relation for $\rmax(E)/\rvir>1$ and the deviation increases with increasing $\rmax(E)/\rvir$. This falloff of the halo data occurs because we only count particles with $r<2.5\rvir$. The halo data also deviate from the scaling relation for $f(E)$ for $\rmax(E)/\rvir < 0.05$, which cannot be attributed to limitation of the simulations: the softening length allows for numerical accuracy down to $r\sim 0.01\rvir$. Instead, such deviations are expected when the density profile is described better by $\rho \propto r^{-1}$ than by the SIS at small radii. For $\rho \propto r^{-1}$, $\Phi(r)-E\propto r-\rmax(E)$ and Eq.~(\ref{eq:eddington}) gives $f(E)\propto \rmax^{-5/2}(E)$ to leading order. Therefore, the scaling relation $f(E) \propto \rmax^{-2.08}(E)$ is too shallow to match the halo data for $\rmax(E)/\rvir < 0.05$. 

Following the same procedure described above, we have also fitted scaling relations of slightly different form:
\begin{align}
\frac{dM}{dE}&=\frac{\alpha\rvir}{G}\left[\frac{\rmax(E)}{\rvir}\right]^m=\frac{0.62}{G\rvir^{0.01}}\rmax^{1.01}(E),\label{eq:dmdermax3}\\
f(E)&=\frac{\beta}{Gv_s\rvir^{2}}\left[\frac{\rmax(E)}{\rvir}\right]^n=\frac{0.031\rvir^{0.08}}{Gv_s}\frac{1}{\rmax^{2.08}(E)}.\label{eq:fermax3}
\end{align}
With $\delta_{\log[(dM/dE)G/\rvir]}=0.019$ and $\delta_{\log[f(E)G\vs\rvir^2]}=0.011$, these alternate scaling relations also provide good description of the halo data, but will not be discussed further in view of their similarity to the scaling relations adopted above.

We have also fitted the scaling relations in Eqs.~(\ref{eq:dmdermax1}) and (\ref{eq:fermax1}) to individual haloes (see Appendix \ref{sec:append}). When uncertainties are taken into account, the sets of $(\alpha,m)$ and $(\beta,n)$ for individual haloes are consistent with those values fitted to the median halo results as shown in Eqs.~(\ref{eq:dmdermax2}) and (\ref{eq:fermax2}). The intrinsic halo-to-halo scatter in these parameters is at the 1--3\% level. For our discussion below, we focus on the scaling relations in Eqs.~(\ref{eq:dmdermax2}) and (\ref{eq:fermax2}).

\section{Application of Scaling Relations to Individual Haloes and Comparison with other Approximations}
\label{sec:use}
The scaling relations for the energy distribution $dM/dE$ and DF $f(E)$ presented in \S\ref{sec:simfit} are obtained from the corresponding median quantities for our halo sample. In this section, we demonstrate that these relations provide an efficient method to model the $dM/dE$ and $f(E)$ for individual haloes and compare them with other approximations. 

\begin{figure*}
\centering
\includegraphics[width=1\textwidth]{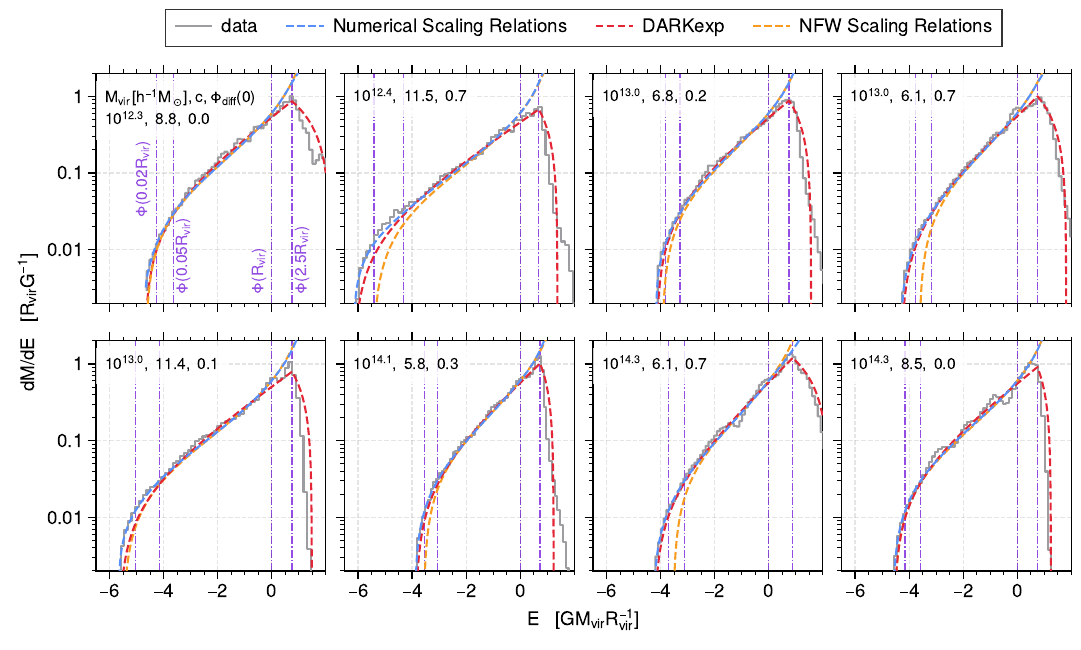}
\vspace{-2em}
\caption{%
Comparison of various approximations to the energy distribution $dM/dE$ with that constructed from simulations (grey curve) for 8 representative isolated and relaxed haloes. The $\rvir$ and $\mvir$ in the units for the axes are taken from the simulations for each halo. In each panel, the blue curve shows the result from the scaling relation in Eq.~(\ref{eq:dmdermax2}) with $\rmax(E)$ obtained from the numerical potential $\Phi(r)$ of each halo, the red curve shows the best-fit DARKexp energy distribution, and the orange curve shows the result from the scaling relation with $\rmax(E)$ obtained from the potential $\Phi_{\rm NFW}(r)$ of the best-fit NFW density profile. The purple vertical lines indicate $E=\Phi(0.02\rvir)$, $\Phi(0.05\rvir)$, $\Phi(\rvir)$, and $\Phi(2.5\rvir)$, respectively. The energy distributions fall off at high energies because only particles inside $2.5R_{\rm vir}$ are counted. The parameter $\Phi_{\text{diff}}(0)=\Phi_{\text{NFW}}(0)-\Phi(0)$ is in units of $G \mvir \rvir^{-1}$. See text for details.
}
\label{fig:dmde}
\end{figure*}

\begin{figure*}
\centering
\includegraphics[width=1\textwidth]{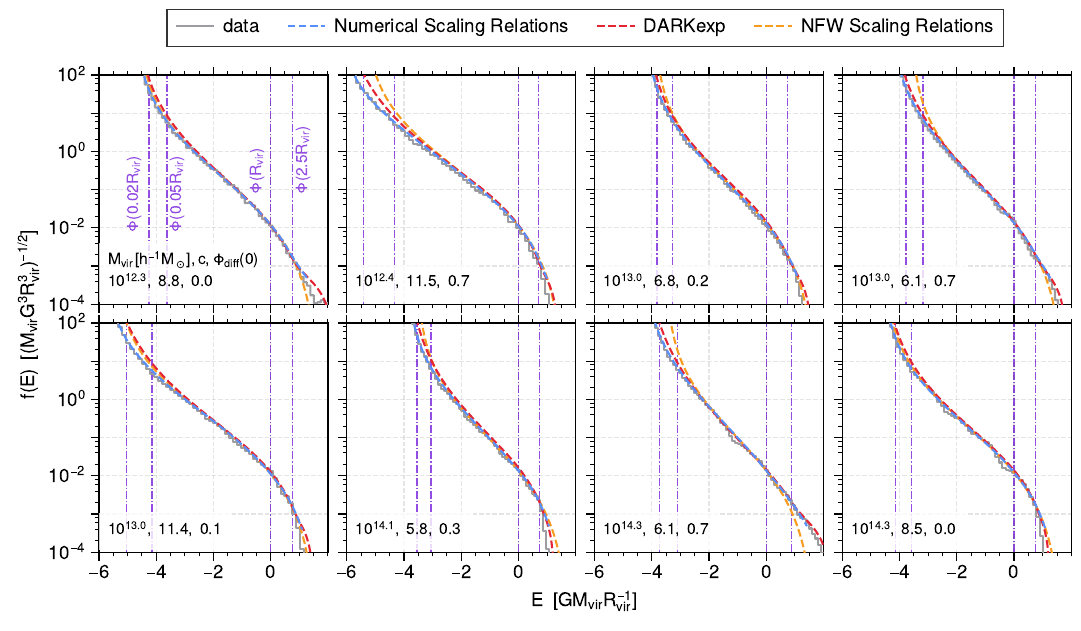}
\vspace{-2em}
\caption{%
Same as Fig.~\ref{fig:dmde}, but for the DF $f(E)$. Note that results from numerical scaling relations use the parameters $\rs$ and $\vs=\rs\sqrt{4\pi G\rhos}$ of the best-fit NFW profile [see Eq.~(\ref{eq:fermax2})]. See text for details.
}
\label{fig:fe}
\end{figure*}

We first show the $dM/dE$ and $f(E)$ constructed from simulations (referred to as ``data'') for 8 representative haloes in Figs.~\ref{fig:dmde} and \ref{fig:fe}, respectively. The $\rvir$ and $\mvir$ in the units for the axes of these figures are taken from the simulations for each halo. For comparison, we also show the $dM/dE$ and $f(E)$ calculated from the scaling relations in Eqs.~(\ref{eq:dmdermax2}) and (\ref{eq:fermax2}), respectively, using the $\rmax(E)$ obtained from the simulated potential for each halo. For convenience, we refer to the latter results as obtained from ``numerical scaling relations'' (simplified to ``scaling relations'' when the context makes the meaning clear) because numerical potentials are used. Note that the $f(E)$ from the numerical scaling relation also uses the parameters $\rs$ and $\vs=\rs\sqrt{4\pi G\rhos}$ of the best-fit NFW profile for each halo. To assess the optimal energy ranges for the scaling relations, we mark the numerical potentials at various radii by vertical lines in Figs.~\ref{fig:dmde} and \ref{fig:fe}. It can be seen that the scaling relations are very good approximations to the data on individual haloes over broad energy ranges [see also Appendix \ref{sec:append}, where the sets of $(\alpha,m)$ and $(\beta,n)$ for the scaling relations fitted to individual haloes are compared with those values used in Eqs.~(\ref{eq:dmdermax2}) and (\ref{eq:fermax2})]. For our entire sample of 79 haloes, we find that the average rms deviations are $\delta_{\log(dM/dE)}=0.04$ over $E\in[\Phi(0.02\rvir),\Phi(\rvir)]$ and $\delta_{\log f(E)}=0.04$ over $E\in[\Phi(0.05\rvir),\Phi(2.5\rvir)]$. These energy ranges correspond to the fitting ranges of the scaling relations. Because the common applicable range of the scaling relations are $E\in[\Phi(0.05\rvir),\Phi(\rvir)]$, we focus on $\delta_{\log(dM/dE)}$ and $\delta_{\log f(E)}$ over this range in comparing various approximations to the $dM/dE$ and $f(E)$ below (see Table~\ref{tab:error}). When appropriate, we also consider these deviations over $E\in[\Phi(0.05\rvir),\Phi(2.5\rvir)]$.

\begin{table*}
	\centering
	\caption{Average rms deviations in $\log(dM/dE)$ and $\log f(E)$ across the entire selected sample of 79 haloes for the numerical scaling relations, DARKexp fits, scaling relations based on the potential of the best-fit NFW profile, and exact results for this profile.}
	\label{tab:error}
    \setlength{\tabcolsep}{3pt} 
    \renewcommand{\arraystretch}{1.2} 
	\begin{tabular}{llcccc} 
		\toprule
            Deviation & Range & Numerical Scaling & DARKexp & NFW Scaling & NFW Exact\\
        \midrule
            $\delta_{\log(dM/dE)}$ & $[\Phi(0.05\rvir),\Phi(\rvir)]$&0.04&0.04&0.06&0.06\\
            $\delta_{\log(dM/dE)}$ & $[\Phi(0.05\rvir),\Phi(2.5\rvir)]$&0.07&0.04&0.08&0.06\\
            $\delta_{\log f(E)}$  &  $[\Phi(0.05\rvir),\Phi(\rvir)]$&0.04&0.07&0.09&0.13\\
            $\delta_{\log f(E)}$  &  $[\Phi(0.05\rvir),\Phi(2.5\rvir)]$&0.04&0.07&0.08&0.12\\
        \bottomrule
	\end{tabular}
\end{table*}

\subsection{Comparison with DARKexp Fits}
\label{sec:darkexp}
Next we compare the quality of fit for the numerical scaling relations and other approximations to the energy distribution and DF. Based on arguments from statistical mechanics, \cite{2010ApJ...722..851H} proposed the DARKexp energy distribution
\begin{equation}
\label{eq:darkexp}
\frac{dM}{dE}=A\left[\exp\left(\frac{E-\Phi_0}{\sigma^2}\right)-1\right],
\end{equation}
where $A$ is a normalization factor, $\Phi_0$ is the central gravitational potential, and $\sigma^2$ is the characteristic energy scale. The above $dM/dE$ and the corresponding density profile have been shown to accurately match those of simulated haloes \citep{Williams_20101,Williams_20102,2015ApJ...811....2H,2016JCAP...09..042N}. 

As our simulated energy distributions are constructed by counting only particles inside $2.5\rvir$, we adopt a modified DARKexp form:
\begin{equation}
\frac{dM}{dE}=
\begin{cases}
\begin{aligned}
        A\left[\exp\left(\frac{E-\Phi_0}{\sigma^2}\right)-1\right], &\ \Phi_0\leq E\leq \Phi(2.5\rvir),\\
        \left(\frac{dM}{dE}\right)_{\Phi_{2.5}}\left(1-\frac{E}{E_\mathrm{max}}\right),&\ \Phi(2.5\rvir)<E\leq E_{\rm max},
\end{aligned}
\end{cases}
\label{eq:darkdmde}
\end{equation}
where $(dM/dE)_{\Phi_{2.5}}$ is the value of $dM/dE$ at $E=\Phi(2.5\rvir)$. In fitting the above form to the simulated $dM/dE$, we first obtain the best-fit parameters $A$ and $\sigma^2$ by taking the central potential $\Phi_0=\Phi(0)$ from the simulations and minimizing the rms difference in $\log(dM/dE)$ over $E\in[\Phi(0.05\rvir),\Phi(2.5\rvir)]$. We then obtain the parameter $E_{\rm max}$ by requiring $\int_{\Phi_0}^{E_{\rm max}}dE(dM/dE)=M_{2.5}$, where $M_{2.5}$ is the mass enclosed within $2.5\rvir$. The best-fit DARKexp energy distributions for the 8 representative haloes are shown in Fig.~\ref{fig:dmde}. For our entire sample of 79 haloes, we find that the average rms deviation is $\delta_{\log(dM/dE)}=0.04$ over $E\in[\Phi(0.05\rvir),\Phi(2.5\rvir)]$, to be compared with 0.07 for the scaling relation in Eq.~(\ref{eq:dmdermax2}). However, both approximations have $\delta_{\log(dM/dE)}=0.04$ over $E\in[\Phi(0.05\rvir),\Phi(\rvir)]$ (see Table~\ref{tab:error}). The worse performance of the scaling relation for $E>\Phi(\rvir)$ is expected because it is fitted over $E\in[\Phi(0.02\rvir),\Phi(\rvir)]$.

To obtain the DF $f(E)$ corresponding to the DARKexp energy distribution in Eq.~(\ref{eq:darkdmde}), we adopt an iterative procedure as in \cite{Gross:2024dsw}. At each iteration, we use the current density profile to determine the associated potential, and then calculate the new density profile from Eqs.~(\ref{eq:densityfromfe}), (\ref{eq:dmde}), (\ref{eq:ge2}), and (\ref{eq:darkdmde}). This procedure is sufficient to obtain the self-consistent density profile $\rho(r)$ for $r\in[0,2.5\rvir]$ and $f(E)$ for $E\in[\Phi_0,E_{\rm max}]$. Our converged density profiles give the correct potential depth $\Phi_{2.5}-\Phi_0$ and enclosed mass $M_{2.5}$. The DARKexp fits to the $f(E)$ are shown in Fig.~\ref{fig:fe} for the 8 representative haloes. For our entire sample of 79 haloes, we find that the average rms deviation is $\delta_{\log f(E)}=0.07$ over $E\in[\Phi(0.05\rvir),\Phi(2.5\rvir)]$, to be compared with 0.04 for the scaling relation in Eq.~(\ref{eq:fermax2}). So the scaling relation is a better approximation to the DF than the DARKexp fit over this energy range.

In general, either the scaling relations or the DARKexp fits can be better approximations to the energy distribution and DF of an individual halo. However, both approximations have comparable quality of fit across the halo sample (see Table~\ref{tab:error}). On the other hand, the DARKexp fits require first fitting the $dM/dE$ to the simulated data, for which the central potential is needed from the simulations, and then calculating the $f(E)$ from an iterative procedure. In comparison, the scaling relations require calculating the $\rmax(E)$ from the simulated potential and obtaining the NFW fit to the simulated density profile. So the scaling relations are easier to apply in practice.

\subsection{Comparison of Results for NFW Profile}
\label{sec:nfw}
As discussed above, applying the numerical scaling relations requires $\rmax(E)$ from the simulated potential and the parameters of the best-fit NFW profile for a halo. It is interesting to examine the results from the scaling relations with $\rmax(E)$ from the potential of the best-fit NFW profile instead. We refer to these results as obtained from ``NFW scaling relations'' and show them for the 8 representative haloes in Figs.~\ref{fig:dmde} and \ref{fig:fe}. For our entire sample of 79 haloes, we find that the average rms deviations are $\delta_{\log(dM/dE)}=0.06$ over $E\in[\Phi(0.05\rvir),\Phi(\rvir)]$ and $\delta_{\log f(E)}=0.08$ over $E\in[\Phi(0.05\rvir),\Phi(2.5\rvir)]$ (see Table~\ref{tab:error}). However, significant deviations occur for $E<\Phi(0.05\rvir)$ when the NFW potential differs significantly from the simulated one at $r<0.05\rvir$ [see haloes with large values of $\Phi_{\rm diff}(0)=\Phi_{\rm NFW}(0)-\Phi(0)$ in Figs.~\ref{fig:dmde} and \ref{fig:fe}]. Because $\rmax(E)$ is calculated from the potential, the difference in the potential between the NFW profile and the simulated halo results in different quality for the fits based on $\rmax(E)$.

As shown in \cite{Gross:2024dsw}, the energy distribution and DF for a simulated halo can be described by those for the best-fit NFW profile to good approximation, but those NFW exact results have the same deficiency as the NFW scaling relations when the NFW potential differs significantly from the simulated one at $r<0.05\rvir$. For our entire sample of 79 haloes, we find that the average rms deviations for the NFW exact results are $\delta_{\log(dM/dE)}=0.06$ over $E\in[\Phi(0.05\rvir),\Phi(\rvir)]$ and $\delta_{\log f(E)}=0.12$ over $E\in[\Phi(0.05\rvir),\Phi(2.5\rvir)]$ (see Table~\ref{tab:error}). So the NFW exact result for the DF is somewhat worse than the NFW scaling relation (see below), while both approximations are comparable in the fit quality of the energy distribution (see Table~\ref{tab:error}).

\begin{figure*}
\centering
\includegraphics[width=1\textwidth]{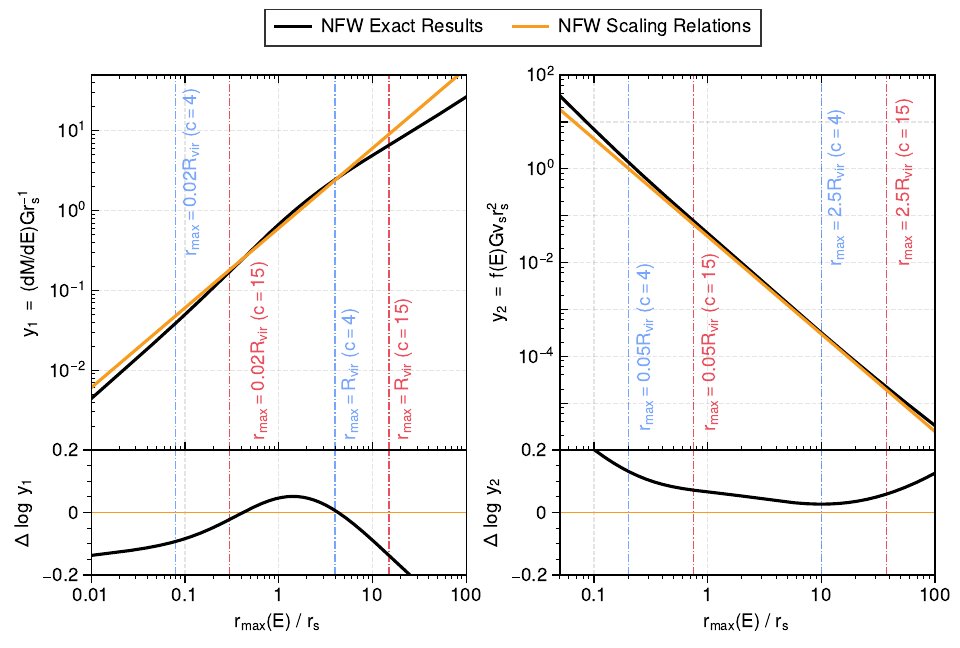}
\vspace{-2em}
\caption{%
\textit{Left panel:} The top part shows $y_1=(dM/dE)G/\rs$ (black curve) as a function of $\rmax(E)/\rs$ for the NFW profile that extends to infinite radius. For comparison, the corresponding scaling relation [Eq.~(\ref{eq:dmdermax2})] is shown as the orange line. The bottom part shows the difference $\Delta\log y_1$ between the black curve and orange line. The blue (red) vertical lines bound the fitting range of $\rmax(E)\in [0.02\rvir, \rvir]$ for the scaling relation for $c=4$ (15) representing the low (high) end of halo concentrations. \textit{Right panel}: Same as left panel, but for $y_2=f(E)Gv_s\rs^2$. The vertical lines bound the fitting range of $\rmax(E)\in [0.05\rvir, 2.5\rvir]$ for the scaling relation in Eq.~(\ref{eq:fermax2}). See text for details.
}
\label{fig:nfw}
\end{figure*}

For a direct comparison of the NFW exact results and scaling relations, we consider a general NFW profile that extends to infinite radius, and show both sets of results in terms of $(dM/dE)G/\rs$ and $f(E)G\vs\rs^2$ as functions of $\rmax(E)/\rs$ in Fig.~\ref{fig:nfw}. The applicable ranges for the NFW scaling relations are $\rmax(E)\in[0.02\rvir,\rvir]$ for $dM/dE$ and $[0.05\rvir,2.5\rvir]$ for $f(E)$, which are indicated by the blue (red) vertical lines for $c=\rvir/\rs=4$ (15) representing the low (high) end of halo concentrations in our sample. When averaged over uniform logarithmic spacing of $\rmax(E)/\rs$ across these ranges, the rms deviations between the NFW exact results and scaling relations are $\delta_{\log[(dM/dE)G/\rs]}=0.05$ (0.05) and $\delta_{\log[f(E)G\vs\rs^2]}=0.07$ (0.04) for $c=4$ (15). In view of such small deviations and considering that the NFW exact results require a rather complicated procedure to obtain the DF from the density profile through Eq.~(\ref{eq:eddington}) while the NFW scaling relations only require inversion of the NFW potential to obtain $\rmax(E)$, we recommend the latter as an efficient method to approximate the energy distributions and DFs of haloes. This recommendation is also supported by the results in Table~\ref{tab:error}. 

As noted above, the NFW scaling relation gives a somewhat better fit to the DF than the NFW exact result (see Table~~\ref{tab:error}). This improvement is mostly due to a better normalization coefficient incorporated in the fit to the simulated data. As shown in the right panel of Fig.~\ref{fig:nfw}, the NFW exact result for the DF lies consistently above the NFW scaling relation although there are only small differences in their shape over the $\rmax(E)$ range of interest.

\section{Connections to Other Halo Models}
\label{sec:origin}
In this section, we make connections between our scaling relations and other halo models. We first discuss the scaling law for the volumetric density of the DF
\begin{align}
\label{eq:vf}
  {\cal{V}}(f_0) = \int d^3{\boldsymbol{r}} d^3{\boldsymbol{v}}
  \delta (f ({\boldsymbol{r}}, {\boldsymbol{v}}) - f_0)
\end{align}
found by \cite{10.1111/j.1365-2966.2004.08045.x}, and then discuss the possible origin of the scaling relation for $dM/dE$.

\subsection{Scaling Law for Volumetric Density of DF}
\cite{10.1111/j.1365-2966.2004.08045.x} found ${\cal{V}}(f) \propto f^{-2.50 \pm 0.05}$ over a wide range of $f$ values for simulated haloes. We now show that this power law can be accounted for by the scaling relations in Eqs.~(\ref{eq:dmdermax2}) and (\ref{eq:fermax2}). For $f({\boldsymbol{r}}, {\boldsymbol{v}})=f(E)$ and $df/dE<0$ (see Fig.~\ref{fig:fe}), we rewrite Eq.~(\ref{eq:vf}) by a change of variable as
\begin{gather}
  {\cal{V}} (f) = -\frac{\int d^3{\boldsymbol{r}} d^3{\boldsymbol{v}}\delta\left(\frac{v^2}{2}+\Phi(r) - E\right)}{df/dE} = -\frac{g(E)}{d f / d E}.
\end{gather}
With $dM/dE\propto\rmax(E)$ [Eq.~(\ref{eq:dmdermax2})] and $f(E)\propto\rmax^{-2.08}(E)$ [Eq.~(\ref{eq:fermax2})], we obtain 
\begin{align}
\label{eq:gevf}
    g(E)=(dM/dE)/f(E)\propto\rmax^{3.08}(E).
\end{align}
From $\Phi(\rmax)=E$, we have
\begin{align}
\label{eq:dedrmax}
    \frac{dE}{d\rmax}=\frac{d\Phi(\rmax)}{d\rmax}=\frac{GM(\rmax)}{\rmax^2(E)},
\end{align}
which gives
\begin{equation}
\label{eq:dfde}
\frac{df}{dE}=\frac{df/d\rmax}{dE/d\rmax}\propto-\frac{\rmax(E)}{M(\rmax)}\frac{1}{\rmax^{2.08}(E)}.
\end{equation}
Combining Eqs.~(\ref{eq:gevf}) and (\ref{eq:dfde}), we obtain
\begin{align}
\label{eq:vf2}
    {\cal{V}}(f)\propto\frac{M(\rmax)}{\rmax(E)}\rmax^{5.16}(E)\propto\frac{M(\rmax)}{\rmax(E)}f^{-2.48}.
\end{align}

The common applicable range for the scaling relations in Eqs.~(\ref{eq:dmdermax2}) and (\ref{eq:fermax2}) is $\rmax(E)\in[0.05\rvir,\rvir]$. For the halo concentration range of $c=4$--15, we can use the result in Eq.~(\ref{eq:vf2}) for $\rmax(E)/\rs\in[0.2,15]$, over which $M(\rmax)/\rmax(E)$ varies within a factor of $\sim 3$ for the NFW profile. In contrast, $f(E)$ varies by a factor of $\sim 10^4$ over the same range of $\rmax(E)/\rs$ (see right panel of Fig.~\ref{fig:nfw}). Therefore, Eq.~(\ref{eq:vf2}) approximately gives ${\cal{V}} (f) \propto f^{-2.48}$, in agreement with the scaling law ${\cal{V}}(f) \propto f^{-2.50 \pm 0.05}$ found by \cite{10.1111/j.1365-2966.2004.08045.x}. 

We note that the scaling relations $dM/dE \propto \rmax(E)$ [Eq.~(\ref{eq:dmdesis})] and $f(E) \propto \rmax^{-2}(E)$ [Eq.~(\ref{eq:fesis})] along with $M(\rmax)\propto\rmax(E)$ for the SIS would give ${\cal{V}}(f) \propto f^{-2.5}$ exactly \citep{10.1111/j.1365-2966.2004.08045.x}. However, Eq.~(\ref{eq:vf2}) provides a better description of ${\cal{V}}(f)$ for the simulated haloes because it expects deviations from an exact power law when $M(\rmax)/\rmax(E)$ varies more strongly with $\rmax(E)$ in the inner and outer regions of haloes. Such deviations were indeed observed by \cite{10.1111/j.1365-2966.2004.08045.x} and more clearly illustrated by \cite{2006MNRAS.373.1293S}.

\subsection{Origin of Scaling Relation for Energy Distribution}
\label{sec:selfsim}
The secondary infall model \citep{1972ApJ...176....1G}, in which dark matter accretes onto an initially collapsed and virialized halo, has long been studied as a model of dark matter halo growth. As first recognized by \cite{1977ApJ...218..592G} and further developed by \cite{1984ApJ...281....1F} and \cite{1985ApJS...58...39B}, infall in this model approximately proceeds in a self-similar manner (see \citealt{10.1111/j.1365-2966.2012.21066.x} for an updated treatment of the model). We now discuss how the scaling relation for $dM/dE$ in Eq.~(\ref{eq:dmdermax2}) may possibly arise during self-similar growth of dark matter haloes. 

For simplicity, we consider a series of mass shells with purely radial motion. Each shell successively reaches its turnaround radius $r_{{\mathrm{ta}}}$ at time $t_{\ast}$ with an enclosed mass of $M_{{\mathrm{ta}}}(t_{\ast})$. Because there is a correspondence between $r_{{\mathrm{ta}}}(t_{\ast})$ and the final energy $E$ of a shell at time $t_0$, or equivalently between $r_{{\mathrm{ta}}}(t_{\ast})$ and $\rmax(E)$ at time $t_0$, we may write
\begin{align}
  \frac{d M}{d E} 
  &= \frac{d M_{{\mathrm{ta}}} (t_{\ast})}{d\ln r_{{\mathrm{ta}}} (t_{\ast})} 
     \frac{d \ln r_{{\mathrm{ta}}} (t_{\ast})}{d \ln \rmax(E)}
     \frac{d \ln \rmax(E)}{d E} \nonumber\\
  &= \frac{d \ln M_{{\mathrm{ta}}} (t_{\ast})}{d \ln r_{{\mathrm{ta}}} (t_{\ast})}
     \frac{d \ln r_{{\mathrm{ta}}} (t_{\ast})}{d \ln \rmax}\frac{M_{{\mathrm{ta}}} (t_{\ast})}{M (\rmax, t_0)}
     \frac{\rmax(E)}{G} \nonumber\\
  &\equiv \kappa \frac{\rmax(E)}{G},
\end{align}
where $d E / d \rmax(E) = {G M (\rmax, t_0)}{\rmax^{-2}}(E)$ is used [see Eq.~(\ref{eq:dedrmax})]. We show below that each factor in the definition of $\kappa$ is approximately constant for typical halo growth, and therefore, the scaling relation in Eq.~(\ref{eq:dmdermax2}) is obtained.

For an initial perturbation with $(\Delta M_i / M_i) \propto M_i^{- \varepsilon}$, where $\varepsilon$ is determined by the index of the matter power spectrum on the relevant scale, a system under self-similar collapse in an Einstein-de Sitter Universe grows as $M_{{\mathrm{ta}}} (t)  \propto t^{2 / 3 \varepsilon}$ and $r_{{\mathrm{ta}}} (t)  \propto M_{{\mathrm{ta}}}^{1 / 3} t^{2 / 3} \propto t^{(2 + 6 \varepsilon) / 9 \varepsilon}$ \citep{2010gfe..book.....M}. So we have $d \ln M_{{\mathrm{ta}}} (t_{\ast})/d \ln r_{{\mathrm{ta}}} (t_{\ast}) = 3/(1 + 3\varepsilon)$. Mass shells initially outside the target shell may move inside it during collapse. Due to this shell crossing and the resulting contraction of particle orbits, $M_{{\mathrm{ta}}} (t_{\ast}) < M (\rmax, t_0)$ and $r_{{\mathrm{ta}}} (t_{\ast}) > \rmax$. If the halo growth rate is low ($\varepsilon > 2 / 3$) or if the accreted matter has non-negligible angular momentum, the inner density profile of a halo will soon settle down \citep[see also \citealt{Zhao2003} for results of cosmological simulations]{1984ApJ...281....1F} so that $M (\rmax, t) / M_{{\mathrm{ta}}}(t_{\ast})$ and $\rmax / r_{{\mathrm{ta}}}(t_{\ast})$ approach constants of the order of unity, and ${d \ln \rmax}/{d \ln r_{{\mathrm{ta}}} (t_{\ast})}\simeq1$. Based on the above discussion, $\kappa$ is approximately a constant. 
We note that the above argument does not hold for fast, purely radial accretion ($\epsilon < 2/3$). However, self-similar halo growth in this case leads to the SIS \citep{2010gfe..book.....M}, for which a scaling relation [Eq.~(\ref{eq:dmdesis})] very close to that in Eq.~(\ref{eq:dmdermax2}) is obtained.

Neither self-similar halo growth with purely radial accretion nor the SIS with a divergent central potential are realistic models of haloes. Nevertheless, they represent reasonable approximations that can shed light on halo structure and formation. The existence of the scaling relation for $dM/dE$ in these simple models suggests that the scaling relations [Eqs.~(\ref{eq:dmdermax2}) and (\ref{eq:fermax2})] presented in this paper may be connected to the universal and fundamental properties imparted to haloes during their formation and relaxation. 
A potential test of this scenario is the dependence of the coefficient $\kappa$ on the halo growth rate. For a higher accretion rate (i.e., a lower $\varepsilon$), we have a larger $d \ln M_{{\mathrm{ta}}} (t_{\ast})/d \ln r_{{\mathrm{ta}}} (t_{\ast})=3/(1 + 3\varepsilon)$. However, this trend is partly cancelled by a smaller $M_{{\mathrm{ta}}} (t_{\ast})/M (\rmax, t_0)$ due to stronger shell crossing of recent accretion (\citealt{1984ApJ...281....1F,Shi2016}). Therefore, the dependence of $\kappa$ on the halo growth rate is weak, which is consistent with the relatively small scatter in the set of $(\alpha,m)$ for the scaling relation fitted to the $dM/dE$ for individual haloes of different growth histories (see Appendix~\ref{sec:append}). This approximate universality makes the scaling relations even more interesting, and also adds to the challenge of understanding their origin.

\section{Summary and Conclusions}
\label{sec:conclude}
We have presented new scaling relations for the energy distributions $dM/dE$ [Eq.~(\ref{eq:dmdermax2})] and DFs $f(E)$ [Eq.~(\ref{eq:fermax2})] of simulated haloes based on the radius $\rmax(E)$ at which the gravitational potential $\Phi(r)$ equals the energy $E$. The proportionality coefficients of these relations depend on the gravitational constant and the parameters $\rs$ and $v_s=\rs\sqrt{4\pi G\rhos}$ of the best-fit NFW profile. Across the entire sample of 79 haloes covering a wide mass range, the average rms deviations for these relations are $\delta_{\log(dM/dE)}=0.04$ over $\rmax(E)\in[0.02\rvir,\rvir]$ and $\delta_{\log f(E)}=0.04$ over $\rmax(E)\in[0.05\rvir,2.5\rvir]$ when the simulated potentials are used. Deviations for other ranges and comparisons with other approximations are given in Table~\ref{tab:error}.

Based on the comparisons in Figs.~\ref{fig:dmde}--\ref{fig:nfw} and Table~\ref{tab:error}, we suggest the following two efficient methods to approximate the energy distributions and DFs of haloes. For simulated haloes, one can use the scaling relations in Eqs.~(\ref{eq:dmdermax2}) and (\ref{eq:fermax2}) after obtaining the best-fit NFW profile and the numerical potential. In more approximate contexts, one can use the potential of the best-fit NFW profile instead of the numerical one. This more approximate method is the most efficient and may be of convenient use in analytical studies of haloes, for which the virial mass $\mvir$ and concentration $c$ can be related to the parameters $\rhos$ and $\rs$ of the NFW profile in a straightforward manner. While the NFW scaling relations have deficiencies at small energy due to significant deviations of the NFW potential from the simulated one (see Figs.~\ref{fig:dmde} and \ref{fig:fe}), these deficiencies may not be relevant in many applications that concern halo regions at $r>0.05\rvir$. Further, the central halo region is affected by baryonic processes, which cannot be addressed by models which only include dark matter. 

Our scaling relations are inspired by those for the SIS and can account for the scaling law for the volumetric density of DF found by \cite{10.1111/j.1365-2966.2004.08045.x}. We have qualitatively discussed how the scaling relation for $dM/dE$ may arise during self-similar halo growth. We also note that the scaling of $f(E)$ with $\rmax(E)$ resembles the scaling law for the pseudo-phase-space density, $\rho(r)/\sigma^3(r) \propto r^{-\gamma}$ \citep{2001ApJ...563..483T}. The true origin of our scaling relations may shed important light on halo formation and merits further study. 

In general, dark matter haloes are only approximately spherical and have velocity anisotropy (see e.g., \citealt{2024ApJ...976..187H}). As mentioned in the introduction, the energy distribution $dM/dE$ of a spherical halo is mostly determined by its density profile and is insensitive to its velocity anisotropy \citep{2008gady.book.....B,2021A&A...653A.140B}. Therefore, so long as spherical density profiles and the corresponding potentials provide good description of the simulated haloes, the scaling relation presented here for $dM/dE$ should also provide good description of these haloes. On the other hand, the DF of the simulated haloes has significant dependence on velocity anisotropy \citep[e.g.,][]{Wojtak2008}. Because the scaling relation presented here for the DF assumes no velocity anisotropy, this result can be applied only when approximation of isotropic haloes is sufficient. While it is challenging to model the energy distribution and DF of anisotropic haloes \citep[but see e.g.,][]{Wojtak2008}, it is worthwhile to investigate how the scaling relations are modified by velocity anisotropy. Such studies may provide insights into the structure and formation of anisotropic haloes, in addition to prescribing good approximations to properties of simulated haloes.

\section*{Acknowledgments}
This work was supported in part by the US Department of Energy under grant DE-FG02-87ER40328 and by a Grant-in-Aid from the University of Minnesota. ZL thanks Jun Zhang for discussion regarding the scaling law of \cite{10.1111/j.1365-2966.2004.08045.x}. ZL acknowledges funding from the European Union’s Horizon 2020 research and innovation programme under the Marie Skłodowska-Curie grant agreement No. 101109759 (``CuspCore'') and from the Israel Science Foundation Grant (ISF 861/20, 3061/21). The figures were created with the colorblind friendly scheme developed by \cite{2021arXiv210702270P}.

\section*{Data Availability}
The data underlying this article will be provided on request to the corresponding author.
\bibliography{draft}

\begin{thebibliography}{}
\makeatletter
\relax
\def\mn@urlcharsother{\let\do\@makeother \do\$\do\&\do\#\do\^\do\_\do\%\do\~}
\def\mn@doi{\begingroup\mn@urlcharsother \@ifnextchar [ {\mn@doi@} {\mn@doi@[]}}
\def\mn@doi@[#1]#2{\def\@tempa{#1}\ifx\@tempa\@empty \href {http://dx.doi.org/#2} {doi:#2}\else \href {http://dx.doi.org/#2} {#1}\fi \endgroup}
\def\mn@eprint#1#2{\mn@eprint@#1:#2::\@nil}
\def\mn@eprint@arXiv#1{\href {http://arxiv.org/abs/#1} {{\tt arXiv:#1}}}
\def\mn@eprint@dblp#1{\href {http://dblp.uni-trier.de/rec/bibtex/#1.xml} {dblp:#1}}
\def\mn@eprint@#1:#2:#3:#4\@nil{\def\@tempa {#1}\def\@tempb {#2}\def\@tempc {#3}\ifx \@tempc \@empty \let \@tempc \@tempb \let \@tempb \@tempa \fi \ifx \@tempb \@empty \def\@tempb {arXiv}\fi \@ifundefined {mn@eprint@\@tempb}{\@tempb:\@tempc}{\expandafter \expandafter \csname mn@eprint@\@tempb\endcsname \expandafter{\@tempc}}}

\bibitem[\protect\citeauthoryear{Arad, Dekel  \& Klypin}{Arad et~al.}{2004}]{10.1111/j.1365-2966.2004.08045.x}
Arad I.,  Dekel A.,   Klypin A.,  2004, \mn@doi [\mnras] {10.1111/j.1365-2966.2004.08045.x}, 353, 15

\bibitem[\protect\citeauthoryear{{Arora} \& {Williams}}{{Arora} \& {Williams}}{2020}]{2020ApJ...893...53A}
{Arora} A.,  {Williams} L. L.~R.,  2020, \mn@doi [\apj] {10.3847/1538-4357/ab7f2e}, \href {https://ui.adsabs.harvard.edu/abs/2020ApJ...893...53A} {893, 53}

\bibitem[\protect\citeauthoryear{{Austin}, {Williams}, {Barnes}, {Babul}  \& {Dalcanton}}{{Austin} et~al.}{2005}]{2005ApJ...634..756A}
{Austin} C.~G.,  {Williams} L. L.~R.,  {Barnes} E.~I.,  {Babul} A.,   {Dalcanton} J.~J.,  2005, \mn@doi [\apj] {10.1086/497133}, \href {https://ui.adsabs.harvard.edu/abs/2005ApJ...634..756A} {634, 756}

\bibitem[\protect\citeauthoryear{{Baes} \& {Dejonghe}}{{Baes} \& {Dejonghe}}{2021}]{2021A&A...653A.140B}
{Baes} M.,  {Dejonghe} H.,  2021, \mn@doi [\aap] {10.1051/0004-6361/202141463}, \href {https://ui.adsabs.harvard.edu/abs/2021A&A...653A.140B} {653, A140}

\bibitem[\protect\citeauthoryear{{Bertschinger}}{{Bertschinger}}{1985}]{1985ApJS...58...39B}
{Bertschinger} E.,  1985, \mn@doi [\apjs] {10.1086/191028}, \href {https://ui.adsabs.harvard.edu/abs/1985ApJS...58...39B} {58, 39}

\bibitem[\protect\citeauthoryear{{Binney} \& {Tremaine}}{{Binney} \& {Tremaine}}{2008}]{2008gady.book.....B}
{Binney} J.,  {Tremaine} S.,  2008, {Galactic Dynamics: Second Edition}.
Princeton University Press

\bibitem[\protect\citeauthoryear{{Bryan} \& {Norman}}{{Bryan} \& {Norman}}{1998}]{1998ApJ...495...80B}
{Bryan} G.~L.,  {Norman} M.~L.,  1998, \mn@doi [\apj] {10.1086/305262}, \href {https://ui.adsabs.harvard.edu/abs/1998ApJ...495...80B} {495, 80}

\bibitem[\protect\citeauthoryear{{Cuddeford}}{{Cuddeford}}{1991}]{1991MNRAS.253..414C}
{Cuddeford} P.,  1991, \mn@doi [\mnras] {10.1093/mnras/253.3.414}, \href {https://ui.adsabs.harvard.edu/abs/1991MNRAS.253..414C} {253, 414}

\bibitem[\protect\citeauthoryear{Eddington}{Eddington}{1916}]{10.1093/mnras/76.7.572}
Eddington A.~S.,  1916, \mn@doi [\mnras] {10.1093/mnras/76.7.572}, 76, 572

\bibitem[\protect\citeauthoryear{Evans \& An}{Evans \& An}{2006}]{PhysRevD.73.023524}
Evans N.~W.,  An J.~H.,  2006, \mn@doi [Phys. Rev. D] {10.1103/PhysRevD.73.023524}, 73, 023524

\bibitem[\protect\citeauthoryear{{Fillmore} \& {Goldreich}}{{Fillmore} \& {Goldreich}}{1984}]{1984ApJ...281....1F}
{Fillmore} J.~A.,  {Goldreich} P.,  1984, \mn@doi [\apj] {10.1086/162070}, \href {https://ui.adsabs.harvard.edu/abs/1984ApJ...281....1F} {281, 1}

\bibitem[\protect\citeauthoryear{{Fong} \& {Han}}{{Fong} \& {Han}}{2021}]{2021MNRAS.503.4250F}
{Fong} M.,  {Han} J.,  2021, \mn@doi [\mnras] {10.1093/mnras/stab259}, \href {https://ui.adsabs.harvard.edu/abs/2021MNRAS.503.4250F} {503, 4250}

\bibitem[\protect\citeauthoryear{{Gao}, {Han}, {Fong}, {Jing}  \& {Li}}{{Gao} et~al.}{2023}]{2023ApJ...953...37G}
{Gao} H.,  {Han} J.,  {Fong} M.,  {Jing} Y.~P.,   {Li} Z.,  2023, \mn@doi [\apj] {10.3847/1538-4357/acdfcd}, \href {https://ui.adsabs.harvard.edu/abs/2023ApJ...953...37G} {953, 37}

\bibitem[\protect\citeauthoryear{{Gross}, {Li}  \& {Qian}}{{Gross} et~al.}{2024}]{Gross:2024dsw}
{Gross} A.,  {Li} Z.,   {Qian} Y.-Z.,  2024, \mn@doi [\mnras] {10.1093/mnras/stae864}, \href {https://ui.adsabs.harvard.edu/abs/2024MNRAS.530..836G} {530, 836}

\bibitem[\protect\citeauthoryear{{Gunn}}{{Gunn}}{1977}]{1977ApJ...218..592G}
{Gunn} J.~E.,  1977, \mn@doi [\apj] {10.1086/155715}, \href {https://ui.adsabs.harvard.edu/abs/1977ApJ...218..592G} {218, 592}

\bibitem[\protect\citeauthoryear{{Gunn} \& {Gott}}{{Gunn} \& {Gott}}{1972}]{1972ApJ...176....1G}
{Gunn} J.~E.,  {Gott} J.~Richard I.,  1972, \mn@doi [\apj] {10.1086/151605}, \href {https://ui.adsabs.harvard.edu/abs/1972ApJ...176....1G} {176, 1}

\bibitem[\protect\citeauthoryear{{Han}, {Jing}, {Wang}  \& {Wang}}{{Han} et~al.}{2012}]{2012MNRAS.427.2437H}
{Han} J.,  {Jing} Y.~P.,  {Wang} H.,   {Wang} W.,  2012, \mn@doi [\mnras] {10.1111/j.1365-2966.2012.22111.x}, \href {https://ui.adsabs.harvard.edu/abs/2012MNRAS.427.2437H} {427, 2437}

\bibitem[\protect\citeauthoryear{{He} et~al.,}{{He} et~al.}{2024}]{2024ApJ...976..187H}
{He} J.,  et~al., 2024, \mn@doi [\apj] {10.3847/1538-4357/ad8882}, \href {https://ui.adsabs.harvard.edu/abs/2024ApJ...976..187H} {976, 187}

\bibitem[\protect\citeauthoryear{{Hjorth} \& {Williams}}{{Hjorth} \& {Williams}}{2010}]{2010ApJ...722..851H}
{Hjorth} J.,  {Williams} L. L.~R.,  2010, \mn@doi [\apj] {10.1088/0004-637X/722/1/851}, \href {https://ui.adsabs.harvard.edu/abs/2010ApJ...722..851H} {722, 851}

\bibitem[\protect\citeauthoryear{{Hjorth}, {Williams}, {Wojtak}  \& {McLaughlin}}{{Hjorth} et~al.}{2015}]{2015ApJ...811....2H}
{Hjorth} J.,  {Williams} L. L.~R.,  {Wojtak} R.,   {McLaughlin} M.,  2015, \mn@doi [\apj] {10.1088/0004-637X/811/1/2}, \href {https://ui.adsabs.harvard.edu/abs/2015ApJ...811....2H} {811, 2}

\bibitem[\protect\citeauthoryear{{Hoffman}, {Romano-D{\'\i}az}, {Shlosman}  \& {Heller}}{{Hoffman} et~al.}{2007}]{2007ApJ...671.1108H}
{Hoffman} Y.,  {Romano-D{\'\i}az} E.,  {Shlosman} I.,   {Heller} C.,  2007, \mn@doi [\apj] {10.1086/523695}, \href {https://ui.adsabs.harvard.edu/abs/2007ApJ...671.1108H} {671, 1108}

\bibitem[\protect\citeauthoryear{{Ludlow}, {Navarro}, {White}, {Boylan-Kolchin}, {Springel}, {Jenkins}  \& {Frenk}}{{Ludlow} et~al.}{2011}]{2011MNRAS.415.3895L}
{Ludlow} A.~D.,  {Navarro} J.~F.,  {White} S. D.~M.,  {Boylan-Kolchin} M.,  {Springel} V.,  {Jenkins} A.,   {Frenk} C.~S.,  2011, \mn@doi [\mnras] {10.1111/j.1365-2966.2011.19008.x}, \href {https://ui.adsabs.harvard.edu/abs/2011MNRAS.415.3895L} {415, 3895}

\bibitem[\protect\citeauthoryear{{Marinacci} et~al.,}{{Marinacci} et~al.}{2018}]{2018MNRAS.480.5113M}
{Marinacci} F.,  et~al., 2018, \mn@doi [\mnras] {10.1093/mnras/sty2206}, \href {https://ui.adsabs.harvard.edu/abs/2018MNRAS.480.5113M} {480, 5113}

\bibitem[\protect\citeauthoryear{{Mo}, {van den Bosch}  \& {White}}{{Mo} et~al.}{2010}]{2010gfe..book.....M}
{Mo} H.,  {van den Bosch} F.~C.,   {White} S.,  2010, {Galaxy Formation and Evolution}.
Cambridge University Press

\bibitem[\protect\citeauthoryear{Naiman et~al.,}{Naiman et~al.}{2018}]{10.1093/mnras/sty618}
Naiman J.~P.,  et~al., 2018, \mn@doi [\mnras] {10.1093/mnras/sty618}, 477, 1206

\bibitem[\protect\citeauthoryear{{Natarajan}, {Hjorth}  \& {van Kampen}}{{Natarajan} et~al.}{1997}]{1997MNRAS.286..329N}
{Natarajan} P.,  {Hjorth} J.,   {van Kampen} E.,  1997, \mn@doi [\mnras] {10.1093/mnras/286.2.329}, \href {https://ui.adsabs.harvard.edu/abs/1997MNRAS.286..329N} {286, 329}

\bibitem[\protect\citeauthoryear{{Navarro}, {Frenk}  \& {White}}{{Navarro} et~al.}{1997}]{1997ApJ...490..493N}
{Navarro} J.~F.,  {Frenk} C.~S.,   {White} S. D.~M.,  1997, \mn@doi [\apj] {10.1086/304888}, \href {https://ui.adsabs.harvard.edu/abs/1997ApJ...490..493N} {490, 493}

\bibitem[\protect\citeauthoryear{Nelson et~al.,}{Nelson et~al.}{2017}]{10.1093/mnras/stx3040}
Nelson D.,  et~al., 2017, \mn@doi [\mnras] {10.1093/mnras/stx3040}, 475, 624

\bibitem[\protect\citeauthoryear{{Nolting}, {Williams}, {Boylan-Kolchin}  \& {Hjorth}}{{Nolting} et~al.}{2016}]{2016JCAP...09..042N}
{Nolting} C.,  {Williams} L. L.~R.,  {Boylan-Kolchin} M.,   {Hjorth} J.,  2016, \mn@doi [\jcap] {10.1088/1475-7516/2016/09/042}, \href {https://ui.adsabs.harvard.edu/abs/2016JCAP...09..042N} {2016, 042}

\bibitem[\protect\citeauthoryear{{Petroff}}{{Petroff}}{2021}]{2021arXiv210702270P}
{Petroff} M.~A.,  2021, \mn@doi [preprint] {10.48550/arXiv.2107.02270}, \href {https://ui.adsabs.harvard.edu/abs/2021arXiv210702270P} {p. arXiv:2107.02270}

\bibitem[\protect\citeauthoryear{Pillepich et~al.,}{Pillepich et~al.}{2017}]{10.1093/mnras/stx3112}
Pillepich A.,  et~al., 2017, \mn@doi [\mnras] {10.1093/mnras/stx3112}, 475, 648

\bibitem[\protect\citeauthoryear{{Planck Collaboration} et~al.,}{{Planck Collaboration} et~al.}{2016}]{2016A&A...594A..13P}
{Planck Collaboration} et~al., 2016, \mn@doi [\aap] {10.1051/0004-6361/201525830}, \href {https://ui.adsabs.harvard.edu/abs/2016A&A...594A..13P} {594, A13}

\bibitem[\protect\citeauthoryear{Posti, Binney, Nipoti  \& Ciotti}{Posti et~al.}{2015}]{10.1093/mnras/stu2608}
Posti L.,  Binney J.,  Nipoti C.,   Ciotti L.,  2015, \mn@doi [\mnras] {10.1093/mnras/stu2608}, 447, 3060

\bibitem[\protect\citeauthoryear{Salvador-Solé, Viñas, Manrique  \& Serra}{Salvador-Solé et~al.}{2012}]{10.1111/j.1365-2966.2012.21066.x}
Salvador-Solé E.,  Viñas J.,  Manrique A.,   Serra S.,  2012, \mn@doi [\mnras] {10.1111/j.1365-2966.2012.21066.x}, 423, 2190

\bibitem[\protect\citeauthoryear{{Schaller} et~al.,}{{Schaller} et~al.}{2015}]{2015MNRAS.451.1247S}
{Schaller} M.,  et~al., 2015, \mn@doi [\mnras] {10.1093/mnras/stv1067}, \href {https://ui.adsabs.harvard.edu/abs/2015MNRAS.451.1247S} {451, 1247}

\bibitem[\protect\citeauthoryear{{Sharma} \& {Steinmetz}}{{Sharma} \& {Steinmetz}}{2006}]{2006MNRAS.373.1293S}
{Sharma} S.,  {Steinmetz} M.,  2006, \mn@doi [\mnras] {10.1111/j.1365-2966.2006.11043.x}, \href {https://ui.adsabs.harvard.edu/abs/2006MNRAS.373.1293S} {373, 1293}

\bibitem[\protect\citeauthoryear{{Shi}}{{Shi}}{2016}]{Shi2016}
{Shi} X.,  2016, \mn@doi [\mnras] {10.1093/mnras/stw925}, \href {https://ui.adsabs.harvard.edu/abs/2016MNRAS.459.3711S} {459, 3711}

\bibitem[\protect\citeauthoryear{Springel et~al.,}{Springel et~al.}{2017}]{10.1093/mnras/stx3304}
Springel V.,  et~al., 2017, \mn@doi [\mnras] {10.1093/mnras/stx3304}, 475, 676

\bibitem[\protect\citeauthoryear{{Taylor} \& {Navarro}}{{Taylor} \& {Navarro}}{2001}]{2001ApJ...563..483T}
{Taylor} J.~E.,  {Navarro} J.~F.,  2001, \mn@doi [\apj] {10.1086/324031}, \href {https://ui.adsabs.harvard.edu/abs/2001ApJ...563..483T} {563, 483}

\bibitem[\protect\citeauthoryear{{Widrow}}{{Widrow}}{2000}]{2000ApJS..131...39W}
{Widrow} L.~M.,  2000, \mn@doi [\apjs] {10.1086/317367}, \href {https://ui.adsabs.harvard.edu/abs/2000ApJS..131...39W} {131, 39}

\bibitem[\protect\citeauthoryear{Williams \& Evans}{Williams \& Evans}{2015}]{10.1093/mnras/stv096}
Williams A.~A.,  Evans N.~W.,  2015, \mn@doi [\mnras] {10.1093/mnras/stv096}, 448, 1360

\bibitem[\protect\citeauthoryear{Williams \& Hjorth}{Williams \& Hjorth}{2010}]{Williams_20101}
Williams L. L.~R.,  Hjorth J.,  2010, \mn@doi [\apj] {10.1088/0004-637x/722/1/856}, 722, 856

\bibitem[\protect\citeauthoryear{Williams, Hjorth  \& Wojtak}{Williams et~al.}{2010}]{Williams_20102}
Williams L. L.~R.,  Hjorth J.,   Wojtak R.,  2010, \mn@doi [\apj] {10.1088/0004-637x/725/1/282}, 725, 282

\bibitem[\protect\citeauthoryear{{Wojtak}, {{\L}okas}, {Mamon}, {Gottl{\"o}ber}, {Klypin}  \& {Hoffman}}{{Wojtak} et~al.}{2008}]{Wojtak2008}
{Wojtak} R.,  {{\L}okas} E.~L.,  {Mamon} G.~A.,  {Gottl{\"o}ber} S.,  {Klypin} A.,   {Hoffman} Y.,  2008, \mn@doi [\mnras] {10.1111/j.1365-2966.2008.13441.x}, \href {https://ui.adsabs.harvard.edu/abs/2008MNRAS.388..815W} {388, 815}

\bibitem[\protect\citeauthoryear{{Zhao}, {Mo}, {Jing}  \& {B{\"o}rner}}{{Zhao} et~al.}{2003}]{Zhao2003}
{Zhao} D.~H.,  {Mo} H.~J.,  {Jing} Y.~P.,   {B{\"o}rner} G.,  2003, \mn@doi [\mnras] {10.1046/j.1365-8711.2003.06135.x}, \href {https://ui.adsabs.harvard.edu/abs/2003MNRAS.339...12Z} {339, 12}

\makeatother
\end{thebibliography}
\bibliographystyle{mnras}
\appendix
\section{Fits for individual haloes}
\label{sec:append}

\begin{figure*}
    \centering
    
    \includegraphics[width=0.85\textwidth]{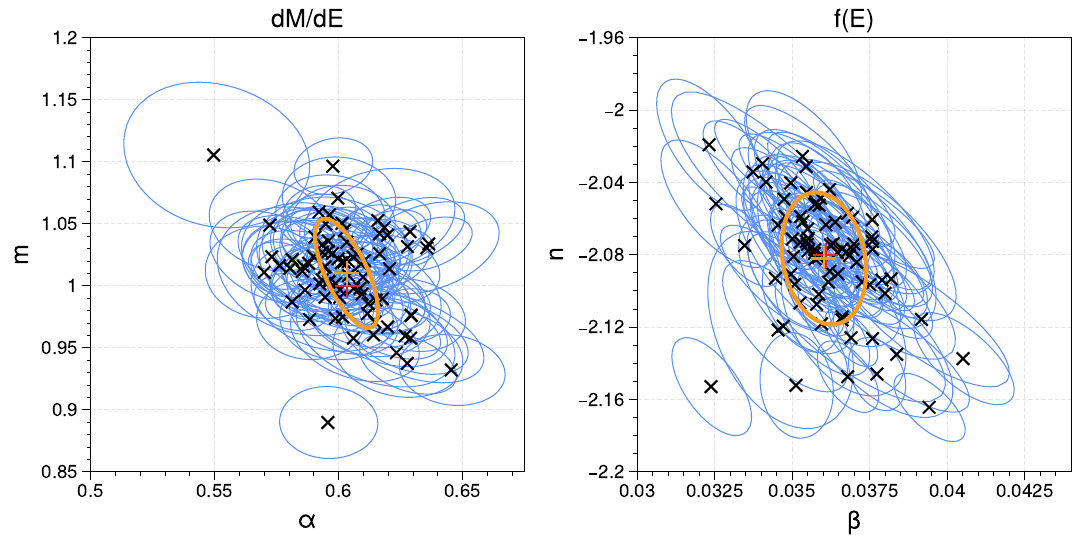}
    \caption{
    \textit{Left panel:} Best-fit parameters $(\alpha, m)$ of Eq.~ (\ref{eq:dmdermax1}) for $dM/dE$ for each of the 79 haloes. 
    \textit{Right panel:} Best-fit parameters $(\beta, n)$ of Eq.~ (\ref{eq:fermax1}) for $f(E)$.
    In both panels, blue ellipses represent the $1\sigma$ (68\%) fitting uncertainties around the best-fit parameters (black crosses) for individual haloes.
    Orange ellipses indicate intrinsic halo-to-halo scatters around the mean best-fit parameters (orange pluses), which nearly coincide with the 
    red pluses for $(\alpha,m)=(0.60,1)$ and $(\beta,n)=(0.036,-2.08)$ obtained from fitting the median halo results in \S\ref{sec:simfit}.
    }
    \vspace{-2em}
    \label{fig:append}
\end{figure*}

We estimate the halo-to-halo scatter of the scaling relations by fitting Eqs.~(\ref{eq:dmdermax1}) and (\ref{eq:fermax1}) to individual haloes using the same procedure described in \S\ref{sec:simfit}. In Fig.~\ref{fig:append}, we present the best-fit sets of $(\alpha, m)$ and $(\beta, n)$ for each of the 79 haloes in our sample as black crosses, along with the $1\sigma$ (68\%) fitting uncertainties as blue ellipses.
The mean best-fit parameters of the sample (orange pluses) are $(\alpha, m)=(0.603, 1.01)$ and $(\beta, n)=(0.036, -2.08)$, nearly identical to those values (red pluses) obtained from fitting the median halo results in \S\ref{sec:simfit}.
The rms fitting uncertainties for individual haloes are $\delta = 0.015$, 0.021, 0.0012, and 0.020 for $\alpha$, $m$, $\beta$, and $n$, respectively.
In comparison, the standard deviations of the best-fit parameters are $\sigma= 0.017$, 0.036, 0.0015, and 0.032, respectively.
We quantify the intrinsic halo-to-halo scatter as $\sqrt{\sigma^2-\delta^2}=0.0084$, 0.029, 0.00090, and 0.024, respectively.
These intrinsic scatters are only at the 1--3\% level relative to the mean best-fit parameters.
We thus conclude that the scaling relations obtained here are nearly universal across a wide range of haloes.

\label{lastpage}
\end{document}